\documentclass[doublecol]{epl2} 
\usepackage{graphicx}
\usepackage{dcolumn}
\usepackage{amsmath}
\usepackage{color}
\usepackage{latexsym}
\usepackage{amssymb}
\usepackage{subfigure}

\title{Heat conduction and phonon localization in disordered harmonic
  crystals} 

\author{Anupam Kundu,\inst{1} Abhishek Chaudhuri,\inst{2}  
  Dibyendu Roy,\inst{3} Abhishek Dhar,\inst{1} 
Joel L. Lebowitz\inst{4} \and Herbert Spohn\inst{5}}

\institute{
      \inst{1} Raman Research Institute - C.V. Raman Avenue, Bangalore 560080, India \\
      \inst{2} Department of Physics and Astronomy - University of Sheffield, Sheffield S3 7RH, United Kingdom \\
      \inst{3} Department of Physics - University of California-San Diego, La Jolla, CA 92093 \\
      \inst{4} Departments of Mathematics and Physics - Rutgers University, Piscataway, NJ 08854\\
      \inst{5} Zentrum Mathematik - Technische Universit\"{a}t M\"{u}nchen, D-85747 Garching, Germany
}


\abstract{
 We investigate the steady state heat current in
 two and three  dimensional isotopically  disordered harmonic lattices.
 Using localization theory as well as kinetic
 theory we estimate the system size dependence of the current. These estimates
 are  compared with  numerical results obtained  using an exact
 formula  for the  current given  in terms of 
a phonon transmission function, as well as by direct nonequilibrium
simulations.  We find that heat conduction by high-frequency modes is 
 suppressed by  localization while  low-frequency modes are strongly
 affected  by boundary conditions. Our {\color{black}heuristic}
arguments show that Fourier's law is valid 
in a three dimensional disordered solid except for special boundary
 conditions.   
We also study the pinned case relevant to localization in quantum
 systems and often used as a model system to study 
the validity of Fourier's law. 
Here we provide the first  numerical
 verification of Fourier's law in three dimensions.
In the two dimensional pinned case we find that localization of phonon modes
leads to a heat insulator.  
}

\begin{document}
\maketitle

\def\bea{\begin{eqnarray}} 
\def\eea{\end{eqnarray}} 
\def\nn{\nonumber}
\def\dg{\dagger}
\def\f{\frac}
\def\la{\langle}
\def\ra{\rangle}
\def\a{\alpha}
\def\b{\beta}
\def\d{\delta}
\def\D{\Delta}
\def\e{\epsilon}
\def\g{\gamma}
\def\k{\kappa}
\def\l{\lambda}
\def\om{\omega}
\def\r{\rho}
\def\s{\sigma}
\def\p{\partial} 
\def\n{\eta}
\def\bn{{\bf{n}}} 
\def\be{{\bf{e}}} 
\def\bk{{\bf{l}}} 
\def\b0{{\bf{0}}} 
\def\etal{{\emph{et~al.}}}
\def\tg{\tilde{\gamma}}
\def\tG{\tilde{\Gamma}}
\def\mG{\mathcal{G}}
\def\mY{\mathcal{Y}}
\def\mX{\mathcal{X}}
\def\mV{\mathcal{V}}
\def\mW{\mathcal{W}}
\def\mU{\mathcal{U}}
\def\ni{\noindent}
\def\ie{\emph{i.e.}}

Transport in random media is a topic of great current interest 
\cite{chang06,schwartz07,hu08,navarro05,billy08,savic08,xu09}. Here we
study the effect of isotopic mass disorder and boundary conditions on heat
conduction and localization in two ($2D$) and three ($3D$) dimensional
harmonic crystals, where the properties of the ordered system are known
exactly. These are systems
of $N^d$ ($d=2,3$) atoms in contact, 
at their surfaces perpendicular to the $x$-axis,  with
heat reservoirs at different temperatures. We shall focus mainly on the
dependence of the heat flux $J$ on $N$. When Fourier's law holds then $J \sim
N^{-1}$ but this is known to be violated in computer simulations \cite{dhar08}
and some experiments \cite{chang08,nika09} in $1D$ and $2D$  systems
where one finds $J \sim N^{-\mu}$ with $\mu \neq 1$.
{\color{black}Earlier work [7,11] on heat conduction in glassy harmonic systems computed
the frequency-dependent thermal diffusivity using the Green-Kubo formula for
different system sizes. However their study did not directly address the
question of  asymptotic size dependence of conductivity.}

We will first describe the precise model studied, then give our
heuristic arguments for calculating the $N$ dependence of $J$  and
finally present the numerical results.

We consider simple cubic lattices  with 
displacements at each lattice site $\bn$  
($n_\nu=1,2,...,N$ for $\nu=1,2,...,d$) given by a scalar variable $x_\bn$.    
In the harmonic approximation the system Hamiltonian  is given by 
\begin{eqnarray}
H &=& \sum_{\bn} \frac{m_{\bn}}{2} \dot{x}_{\bn}^2 
+\sum_{n_1=1}^{N-1} \sum_{\bn',\hat{\be}}
\frac{k}{2}(x_{\bn}-x_{\bn+\hat{\be}})^2 \nn \\
&+&\sum_{\bn'} \frac{k'}{2}(x_{1,\bn'}^2+x_{N,\bn'}^2) +\sum_\bn
\frac{k_o}{2} x_\bn^2~\label{ham}
\end{eqnarray}
where $\hat{\be}$ refers to the $d$ unit vectors and we have denoted
$\bn=(n_1,\bn')$. We impose periodic boundary conditions (BCs) in the
directions $\nu=2,3,...,{\color{black}d}$ and consider two different BCs in the
direction of heat conduction ($\nu=1$): (i) fixed BCs
$k' >0 $ and (ii) free BCs $k' =0$.
Pinning refers to the case with $k_o > 0$. 
We consider  binary mass disorder with equal number of particles of 
masses $\bar{m}-\Delta$ and  $\bar{m}+\Delta$ distributed randomly on the
lattice sites. The equations of motion of particles in
the bulk ( $1<n_1<N$ ) are given by: 
$ {m}_{\bn} \ddot{x}_{\bn} = -\sum_{\pm\hat{\be}}k (x_\bn-x_{\bn+\hat{\be}}) -k_o
x_{\bn} $.
The particles at the surfaces $n_1=1$ and $n_1=N$ are connected  
to heat reservoirs, at temperatures $T_L$ and $T_R$ respectively.
These  are modeled by white noise Langevin equations.
This means that  particles at  $n_1=1$ and $n_1=N$, have
additional forces given respectively by  
 $(-k'x_\bn-\gamma \dot{x}_{\bn}+\eta^L_{\bn})$ and $(-k'x_\bn-\gamma
\dot{x}_{\bn}+\eta^R_{\bn})$, where $\eta^{L}$ and $\eta^R$ are noise
terms with
strengths proportional to  $T_L$ and $T_R$ respectively  and to the friction
constant $\gamma$. 
If we model the heat reservoirs  themselves  by  infinite
ordered harmonic crystals then  Langevin type equations
for the system \cite{dharroy06} are obtained on
eliminating the bath degrees of freedom. 
The  two different BCs then emerge naturally.
Fixed BCs  correspond to reservoirs   with  properties different from 
the system ({\emph{e.g.}}   different spring constants) 
while free BCs   correspond to   the case where  the
reservoir is simply an  extension of the system (without   disorder) 
 \cite{rubin71,savic08,stoltz08}. We note that fixed BCs are more
 realistic.

In the nonequilibrium steady state, the {\color{black}heat current per unit area (assuming unit lattice spacing length)} 
from the left to the right reservoir is given by {\color{black}\cite{dharroy06,casher71}}:
\bea
\mathcal{J} &=& \f{\Delta T } {2\pi N^{d-1}} \int_{0}^\infty  d \om {\mathcal{T}}
(\om)~, \label{jeq}   
\eea
where $\Delta T=T_L-T_R$  
 and $\mathcal{T}(\omega)$ is the transmission
coefficient of phonons at frequency $\om$ from the left to the right
reservoir. Our interest here are in the disorder averaged transmission
$T(\om)=[ \mathcal{T}(\om)]/N^{d-1}$ and the current $J=[\mathcal{J}]$.

The phonon  and electron localization problems are
closely related. 
For the system without baths consider the displacement field
$a_{\bn}(p)$ for the $p^{\rm th}$ 
normal mode. This satisfies the equation:
$ m_{\bn} \om_p^2 a_{\bn}=  (2d +k_o) a_{\bn}- \sum_{\hat{\be} }
a_{\bn+\hat{\be}}$. 
By introducing variables $\psi_{\bn}(p)=m_{\bn}^{1/2} 
a_{\bn}(p)$, $v_\bn=(2d  +k_o)/m_\bn$ and $t_{\bn,\bk}= 1/(m_\bn
m_\bk)^{1/2}$ for nearest neighbour sites $\bn,\bk$ this
 equation transforms to a Schrodinger-type equation 
$\omega_p^2 \psi_{\bn}(p)=v_\bn \psi_{\bn} (p) - \sum_\bk t_{\bn,\bk}
\psi_\bk (p)$~. The nearest neighbour hopping 
$t_{\bn,\bk}$ and on-site energies $v_\bn$ are 
now correlated random variables.  For the unpinned case with $k_o=0$,
translation invariance gives rise to  extended modes at low frequencies. The
pinned case has no translational invariance and is closer to the
usual electron localization problem. 

The effect of disorder on heat conduction in a harmonic crystal manifests
itself in two ways: (i) Anderson localization \cite{anderson58} of phonon
modes of frequency $\om$ will make them non-conducting, (ii) non-localized
phonons will be scattered by the impurities.
From Eq.~(\ref{jeq}) the net current is given by the integrated
transmission and for large system sizes the integral is over the range
of normal modes. 
It is natural to classify normal modes 
as localized, diffusive or ballistic. 
For modes localized on a length scale
$\ell$,   $T(\om)\sim e^{-N/\ell}$. This $\ell$ depends on the
phonon frequency and low frequency modes for which $\ell(\om)
\stackrel{>}{\sim} N$ 
will therefore be  carriers of  the heat current. Diffusive modes are
spatially extended but non-periodic and 
$T(\om) \sim 1/N$. Ballistic modes are 
extended and approximately periodic and their transmission is
$N$-independent. 

A renormalization group study in a disordered continuum elastic model by
John {\emph{etal}} \cite{john83} found that in 1D and 2D all non-zero frequency phonons
are localized. They studied the spreading of an energy pulse to define a
frequency dependent diffusivity $D_0({\omega})$. From the behaviour of
$D_0({\omega})$ under renormalization one can obtain a differential recursion
relation for the resistivity. This relation shows that in the large system
size limit the RG flow is towards infinity for dimensions $\le 2$ in contrast
to $d=3$ where the flow is towards zero as long as $\omega$ is less than some
fixed value, independent of system size. Hence all finite-frequency modes in
one and two dimensions are localized. From the differential recursion
relations one finds that for $d=1$ and $d=2$ the localization length in
$\omega \rightarrow 0$ limit diverges as $\sim ~ 1/\omega^2$ and
$\sim~e^{1/\omega^2}$  respectively. In $3D$ there exists a frequency, independent of system size, above which all states are localized while 
states below that frequency are extended. Hence for a system of size $N$ there will be a cut-off frequency $\omega^L_c$ (which depends on $N$ for $d \le 2$) above which all the modes are localized. In different dimensions $\omega^L_c$ is given by
\bea
\omega^L_c &\sim& N^{-1/2}~~{\rm{for}}~d=1 \nonumber \\
&\sim& [log(N)]^{-1/2}~~{\rm{for}}~d=2  \\
&\sim&{\rm{nonzero~value~independent~of}}~N~{\rm{for~}}d=3 \nonumber 
\eea
\noindent
For extended modes with $\om < \om_c^L$ we use kinetic
theory to determine the cut-off frequency $\om_c^K$ below which states are
ballistic. 
Rayleigh scattering of phonons 
gives a mean free path $\ell_K(\omega) \sim \omega^{-(d+1)}$ in
$d$-dimensions. This gives 
$ \om_c^K= N^{-1/(d+1)}$ below which 
$\ell_K(\om)  \stackrel{>}{\sim}  N $.

The net current in the  system consists of both ballistic  
[$J_B \sim \int_0^{\om_c^K} d\om T(\om)$] and diffusive [$J_D \sim
\int_{\om_c^K}^{\om_c^L} d \om T(\om)$) contributions, $J=J_{B}+J_D$.   
One other
crucial observation, consistent with the numerics, is that  $T(\om)$
in the ballistic region has the same form as for an
ordered system, which  is sensitive to boundary conditions.

In $1D$ both $\om_c^K, \om_c^L \sim N^{-1/2}$ and numerical studies 
show that all extended states
are ballistic \cite{dhar01}. For free BC  $T(\om)\to {\rm const.}$ as $\om 
\to 0$ and  hence $J \sim \int_0^{N^{-1/2}} d \om \sim N^{-1/2}$. 
For fixed BC  $T(\om) \sim \om^2$ leading to $J \sim
N^{-3/2}$. 
The dependence of the form of $T(\omega)$ on pinning was discussed in \cite{droy08} in detail.
Thus surprisingly one finds a strong dependence on BCs. In the
presence of pinning, the low frequency modes are 
removed and $J \sim e^{-c N}$. These
results  agree with earlier rigorous and numerical work on this system
\cite{matsuda70,rubin71,casher71,dhar01}.        
We now calculate the asymptotic system size dependence in $2D$ and $3D$.

\begin{figure}
\includegraphics[width=3.2in,height=1.8in]{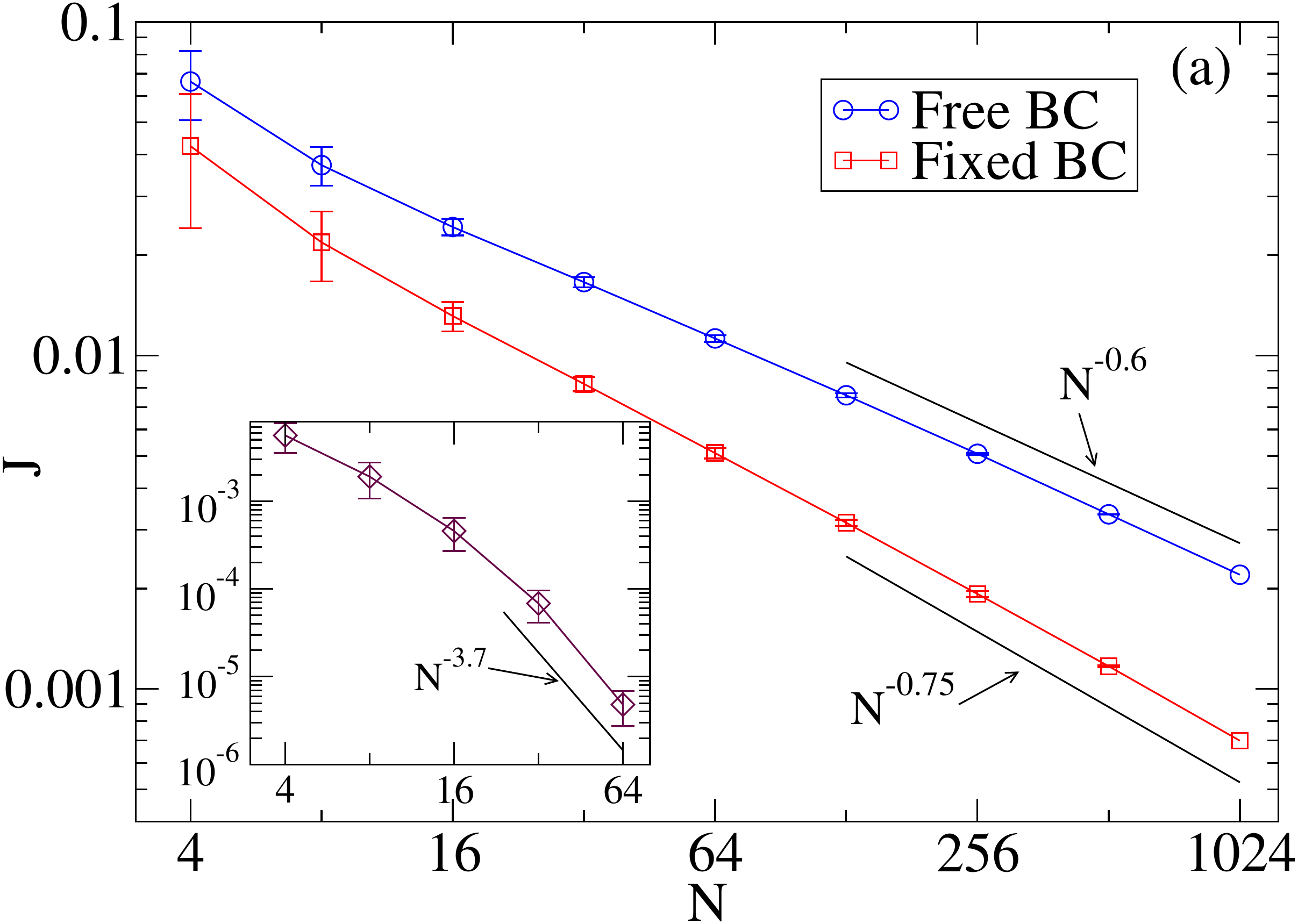} 
\vspace{0.2in} 

\includegraphics[width=3.2in,height=1.8in]{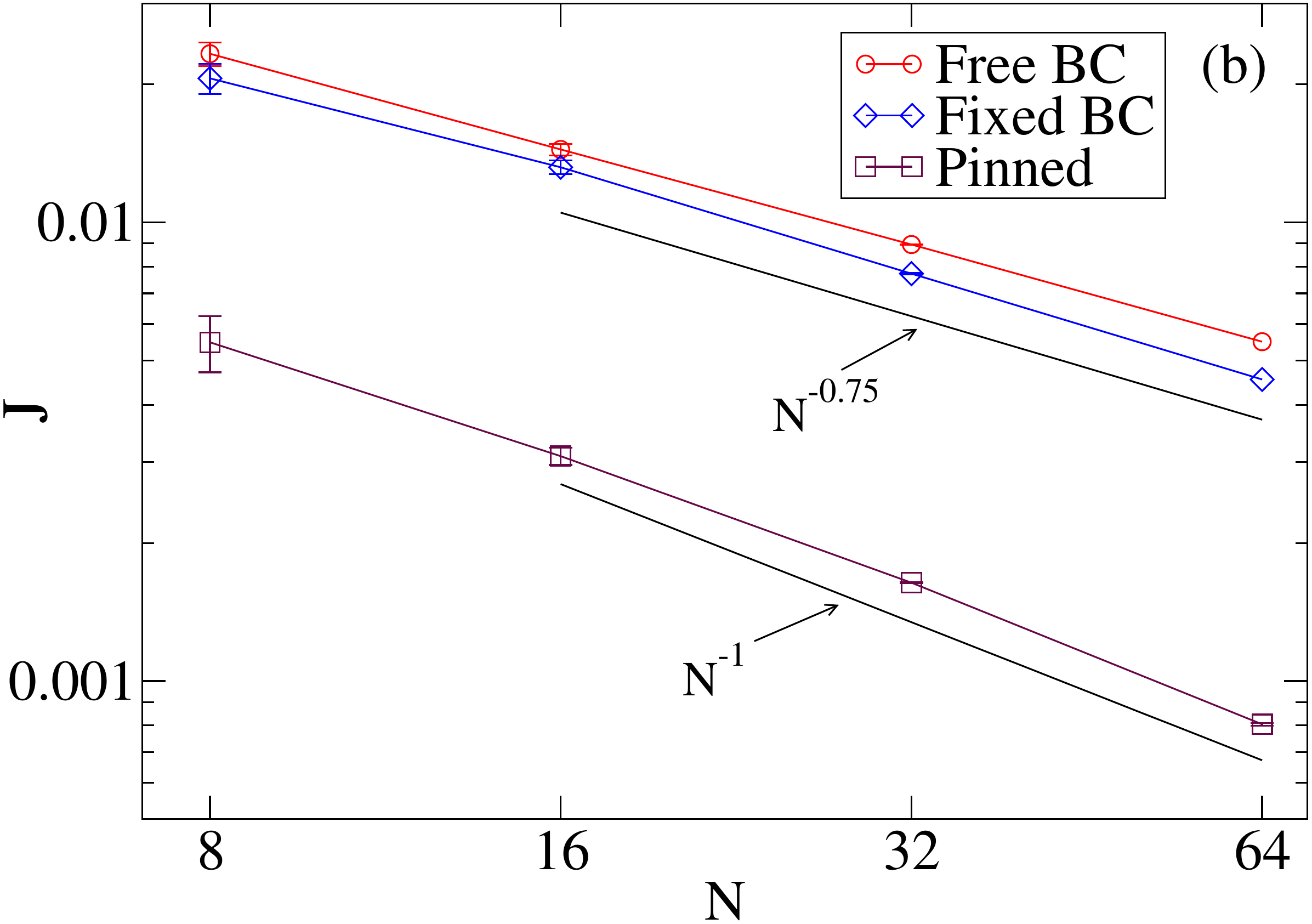}
\caption[Optional caption for list of figures]{
Plot of the $N$-dependence of disorder averaged current $J$. (a) $2D$: Inset
shows result for pinned case. For free BC $\Delta=0.8$ and for fixed BC
$\Delta=0.95$. For pinned case $\Delta=0.4, k_o=10.0$. 
(b) $3D$: For all cases $\Delta=0.8$ and for pinned case $k_o=10.0$. Error
bars show standard deviations due to sample-to-sample fluctuations and are very
small except in the $2D$ pinned case.    }
\label{jvsn}
\end{figure}

\begin{table}
\scalebox{0.8}{%
\begin{tabular}{|c|c|c|c|c|}
\hline
\multicolumn{1}{|c|}{} &
\multicolumn{2}{|c|}{$d=2$} & 
\multicolumn{2}{|c|}{$d=3$}  \\
\cline{2-5}
 & {\color{black}Heuristic} & Numerical & {\color{black}Heuristic} & Numerical  \\ 
\hline 
Pinned & $\exp{(-b N)} $ & $N^{-3.7}$ & $N^{-1}$ & $N^{-1.0}$ 
  \\ 
\hline 
Fixed & $N^{-1}(\ln{N})^{-1/2} $ & $N^{-0.75}$ & $N^{-1}$ & $N^{-0.75}$   \\
\hline 
Free & $N^{-2/3}$ & $N^{-0.6}$ & $N^{-3/4}$ & $N^{-0.71}$   \\
\hline
\end{tabular}}
\caption{The table  gives the $N$ dependence of $J$ and summarizes the main
  results of the paper.  
 The error bar for  the numerically obtained exponent values is  of the order
 $ \pm 0.02$.
N.B: The system sizes used may be far from asymptotic.} 
\label{table}
\end{table}

 {\it Unpinned lattice}: To find the ballistic contribution we note that for the ordered lattice $T(\omega) \sim \omega^{d-1}$ as $\omega \rightarrow 0$ for free BCs and as $T(\omega) \sim \omega^{d+1}$ for
fixed BC. This can be seen in the following way. We write 
${\bf{n}}=(n_1,{\bf{n}}')$ and ${\bf{q}}=(q_2,q_3,...,q_{d})$ with
$q_{\alpha}=\frac{2\pi n}{N}$ where $n$ goes from $1$ to $N$. Now if we define 
$x_{n_1}({\bf{q}}) = N^{-{(d-1)/2}}
\sum_{{\bf{n}}'}~x_{(n_1,{\bf{n}}')}~e^{i{\bf{q}}.{\bf{n}}'}$, 
then one can show that,  for each  ${\bf{q}}$, $x_{l_1}({\bf{q}})$ satisfies a
Langevin equation corresponding to a one-dimensional Hamiltonian with the onsite
spring constant $k_o$ replaced by 
$ \lambda({\bf{q}})=k_o + 2k(d-1 - \sum_{\alpha=2,d} cos(q_{\alpha}))$.
This means that the problem of heat
conduction in a $d$-dimensional ordered harmonic lattice can be related to heat
conduction across $N^{d-1}$ independent ordered harmonic chains with different
onsite potentials \cite{Nakazawa68, Nakazawa70, droyord}. Hence the transmission coefficient $T(\omega)$ for
the $d$-dimensional lattice can be expressed as a sum of the transmission
coefficients of the $1D$ ordered chains. Using this result and the analytic
form of the transmission coefficient for the $1D$ chain \cite{droyord} we find in the $N
\rightarrow \infty$  and $\omega \rightarrow 0$ limit 
\bea
T(\omega) &\sim & \omega^{d-1}~~{\rm{for~open~boundary~condition}} \nonumber \\
&\sim& \omega^{d+1}~~{\rm{for~fixed~boundary~condition}}. \nonumber
\eea

\noindent 
Hence we  get the ballistic 
contribution  to the total current density (for the unpinned case) as:
$J_{B } \sim  \int_0^{\om_c^K} d \om~ \om^{d-1} \sim
{ N^{-d/(d+1)}}$ for free BC  and $ J_{B} \sim  \int_0^{\om_c^K} d
\om~ \om^{d+1} \sim {N^{-(d+2)/(d+1)}}$ for   fixed~~BC. 
In $2D$  using kinetic theory and localization theory we  expect
 localized modes for $\om \stackrel{>}{\sim} \om_c^L=(\ln N)^{-1/2}$,
 ballistic modes for    $\om \stackrel{<}{\sim} \om_c^K= N^{-1/3}$ and  
 diffusive modes for $ \om_c^K \stackrel{<}{\sim} 
 \om \stackrel{<}{\sim} \om_c^L $.
The diffusive contribution to total current
 will  scale as $J_{D} \sim (\ln N)^{-1/2} N^{-1}$. 
As argued above, the  ballistic contribution depends on BCs with 
 $J_{B} \sim N^{-4/3}$ for fixed BC and $J_{B} \sim N^{-2/3}$ for free
BC. Hence,  
 adding all the contributions, we conclude that
 asymptotically:
$J \sim    (\ln N)^{-1/2} N^{-1}$ for  fixed BC and
$J \sim    {N^{-2/3}}$ for free BC.
In $3D$ we expect that $\om_c^L$ is independent of $N$ and states with
$\om > \om_c^L$ are localized.
Extended modes with $\om \stackrel{<}{\sim} \om_c^K= N^{-1/4}$ are
ballistic while those with  
 $ \om_c^K \stackrel{<}{\sim} \om \stackrel{<}{\sim} \om_c^L$ are diffusive.
The contribution from the diffusive modes  scales as $J_{D} \sim
N^{-1}$ while the  
ballistic contribution again depends on boundary conditions 
with  $J_{B} \sim N^{-5/4}$ for fixed BC and $J_{B} \sim
N^{-3/4}$ for free BC. We conclude that asymptotically:
$ J \sim    {N}^{-1}$  for 
  fixed BC and 
$ J \sim  { N^{-3/4}}$ 
for free BC.

\begin{figure*}[t!]
\hspace{0cm}
\subfigure[Unpinned $2D$ lattice]{
\includegraphics[width=3.2in]{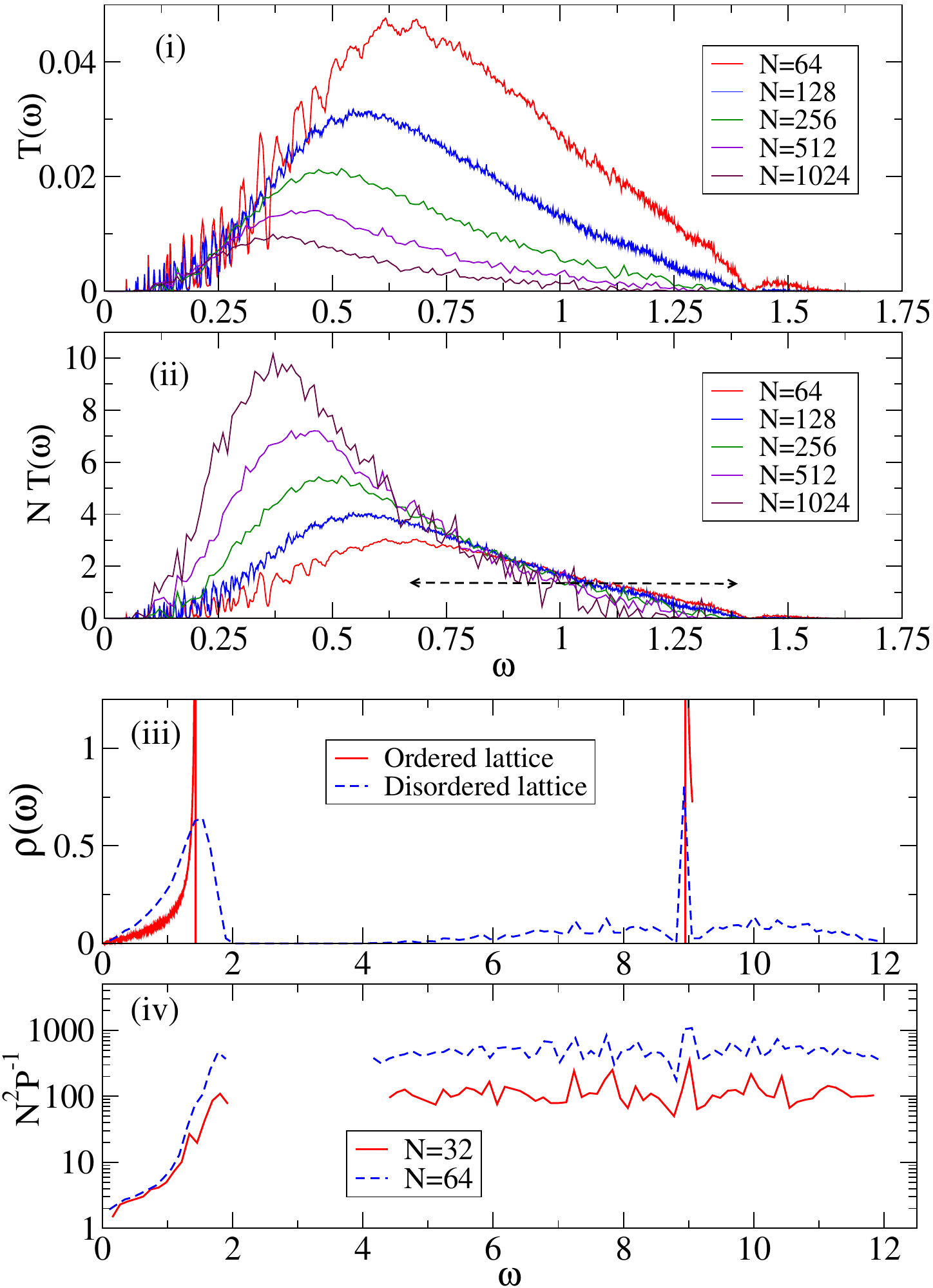}
\label{twipr2dfixed0.95}
\vspace{.25cm}
}
\subfigure[Unpinned $3D$ lattice]{
\includegraphics[width=3.2in]{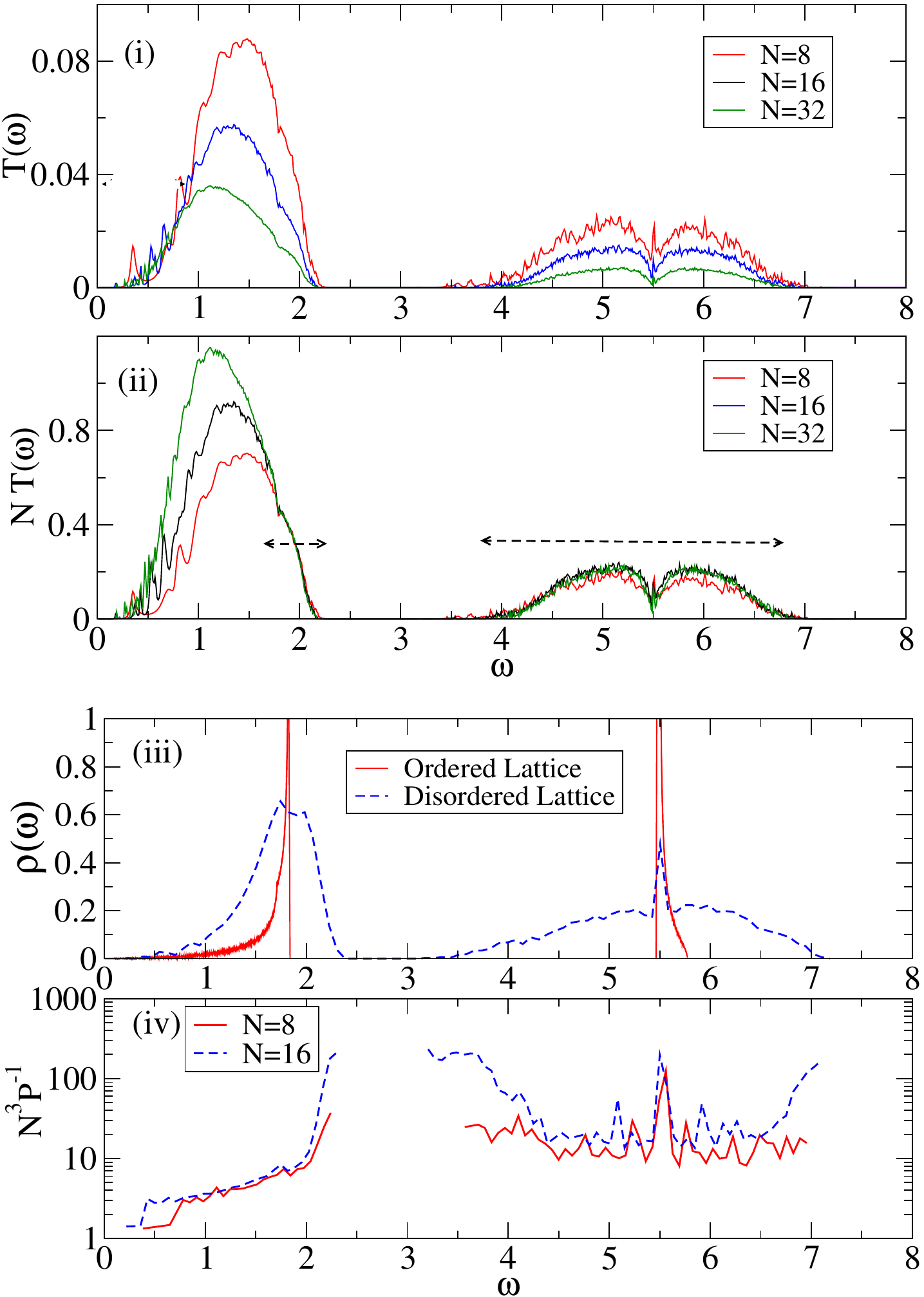}
\label{twipr3dfixed0.8}
\vspace{.25cm}
}
\caption[Optional caption for list of figures]{
 (i) Plot of 
  $T(\om)$.  (ii)   Plot  of $N T(\om)$. The range of frequencies for which
 $T(\om)\sim 1/N$  is indicated  by arrowed lines and corresponds to
  diffusive modes  . (iii) Plot
  of $\rho(\om)$  for   binary mass ordered lattice and  single disordered sample.  (iv)
  Plot of $N^{d-1} \times P^{-1}$ for single  samples. 
Non-collapse of plots for   different $N$ in (iv) indicate localized
 states.  }
\label{twplots}
\end{figure*} 

 {\it Pinned lattice}: There is now a gap in the spectrum, starting from $\om=0$, and thus there
are no low frequency ballistic modes. In $2D$ localization theory then tells us
that for sufficiently large $N$, all non-zero 
frequency modes become localized, hence we should get an insulator. In $3D$
there is a finite band of diffusive states and therefore we expect a  normal
conductor satisfying Fourier's law.   

{\it Numerical results:}
We now check the above predictions through extensive numerical
calculations. 
We measure 
force-constants in units of $k$,  masses in
units of the average mass $\bar{m}$, time  in units of the inverse
frequency $\Omega^{-1} = (\bar{m}/k)^{1/2}$, displacements in units of the
lattice spacing $a$, friction constant $\gamma$ in units of $\bar{m}
\Omega$,   and temperature   in units of $\bar{m} a^2
\Omega^2/k_B$. We set $k=1, \bar{m}=1$ and we fixed $\Delta T=1,
\gamma=1$. For fixed BCs and pinned cases $k'=1$. Different values of the mass 
variance $\Delta$ and the on-site spring constant $k_o$ were studied
for $2D$ and $3D$ lattices of different sizes \cite{abhishek09}. Here we
report results for 
cases with the strongest disorder.  
The transmission coefficient $\mathcal{T}_N(\om)$ can
be expressed in terms of 
phonon Green's functions and can be accurately determined numerically
using a transfer matrix representation \cite{abhishek09}.
By performing a discrete sum over the transmitting range of frequencies we
evaluate  the integration in 
Eq.~(\ref{jeq})  to obtain   $\mathcal{J}$. 
For $N \geq 64$ in $3D$ the transfer
matrix method has numerical problems and in those cases we  performed
nonequilibrium simulations to find the heat current.

We also studied the properties of the normal modes of the
isolated harmonic lattices with fixed BCs. 
We measure  the degree of localization
of a given mode by the inverse participation ratio defined as
$P^{-1}= { \sum_\bn a_\bn^4 }/{(\sum_\bn a_\bn^2)^2}$.
 For an extended  state  $P^{-1}$ is of order $\mathcal{O}(N^{-d})$  while
 for a  localized state, it is  $\mathcal{O}(1)$.

In Fig.~(\ref{jvsn}) we show the $N$-dependence of the disorder averaged
current $J=[\mathcal{J}]$ in $d=2,3$.
Our findings and the comparison
with the {\color{black}heuristic} predictions are summarized
in  Table~(\ref{table}). 
The results for the
free BC case are consistent with the {\color{black}heuristic} predictions
while those for fixed BC show significant deviations.
 For the pinned system we obtain, in agreement with the
theoretical prediction. In 2D we find a strong decay of the current with $N$ with $\mu=3.7$ suggesting that at larger system sizes we will get an exponential decay corresponding to a heat insulator. In 3D we find $\mu=1$ implying that the system is a normal heat conductor. For the binary mass distribution we do not find a transition to insulating behaviour with increasing disorder.

The $N$-dependence of the  disorder-averaged phonon transmission coefficient
$T(\om)=[\mathcal{T}_N(\om)]/N^{d-1}$ sheds additional light on the nature of
phonons in different frequency regions.
For the case of fixed BCs
we show in Fig.~(\ref{twplots})  results for 
transmission, density of states and inverse participation ratios and from
these we can see the range of allowed modes and their degree of localization. 
By plotting $NT(\om)$ we identify the diffusive regime.
From plots (iii-iv) we see that  
in both $2D$ and $3D$ there are effectively two phonon bands, a
remnant of the ordered binary mass case. In $2D$  the upper
band is  fully localized while in $3D$ there is a small number of
localized states near the  band edges.  In $2D$ the lower band 
has extended states below a cut-off $\om_c^L$ which decreases slowly
 with $N$.  In both $2D$ and $3D$ the lower band has diffusive and
 ballistic states and the 
 crossover scale  $\om_c^K$  decreases with system size.  
The expected $N-$dependence of $\om_c^K$ and $\om_c^L$ are difficult 
to verify at these system sizes.

\begin{figure}[t!]
\subfigure[$2D$ lattice ]{
\includegraphics[width=3.2in,height=1.7in]{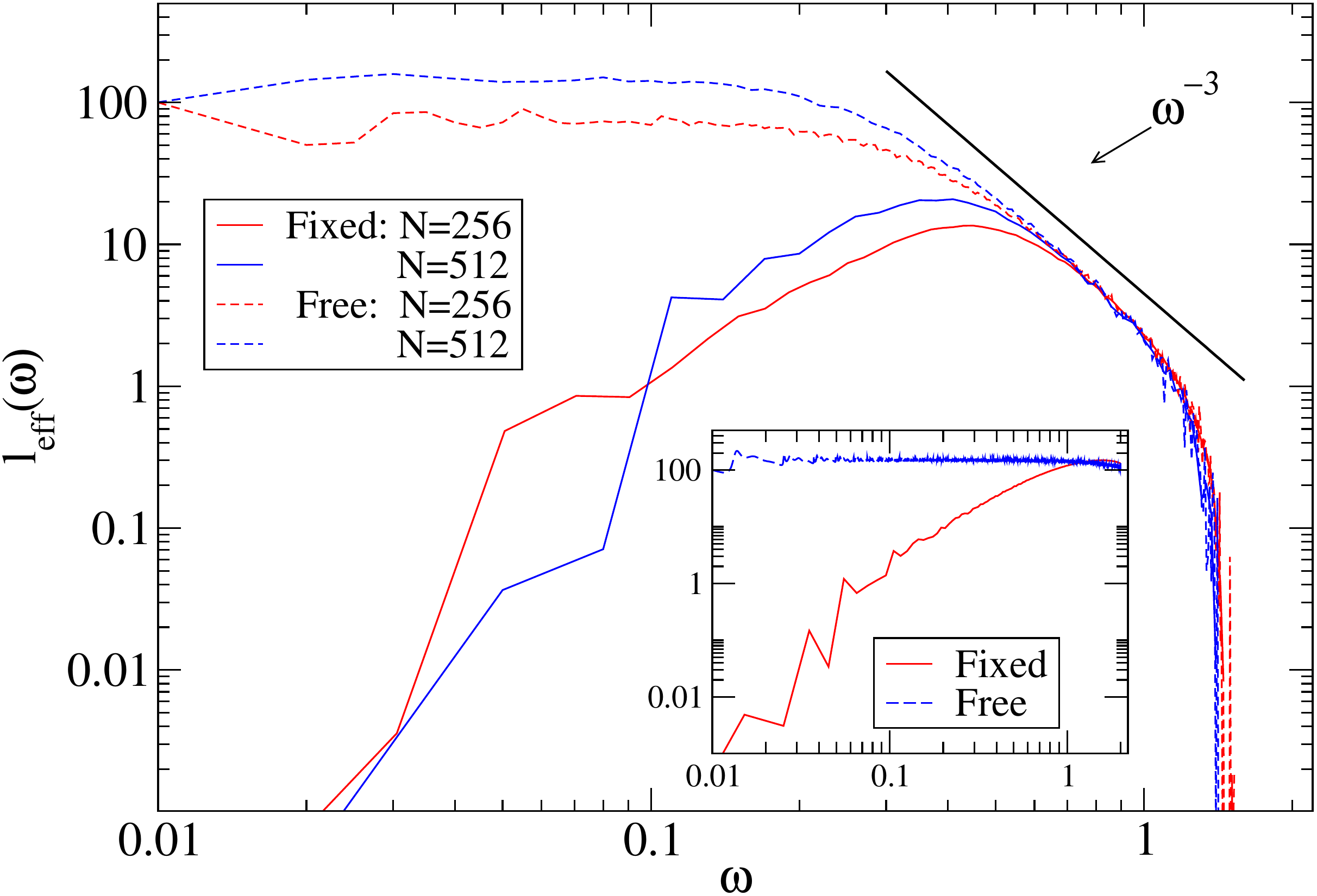}
\label{fig:subfig1}
}
\subfigure[$3D$ lattice]{
\includegraphics[width=3.2in,height=1.7in]{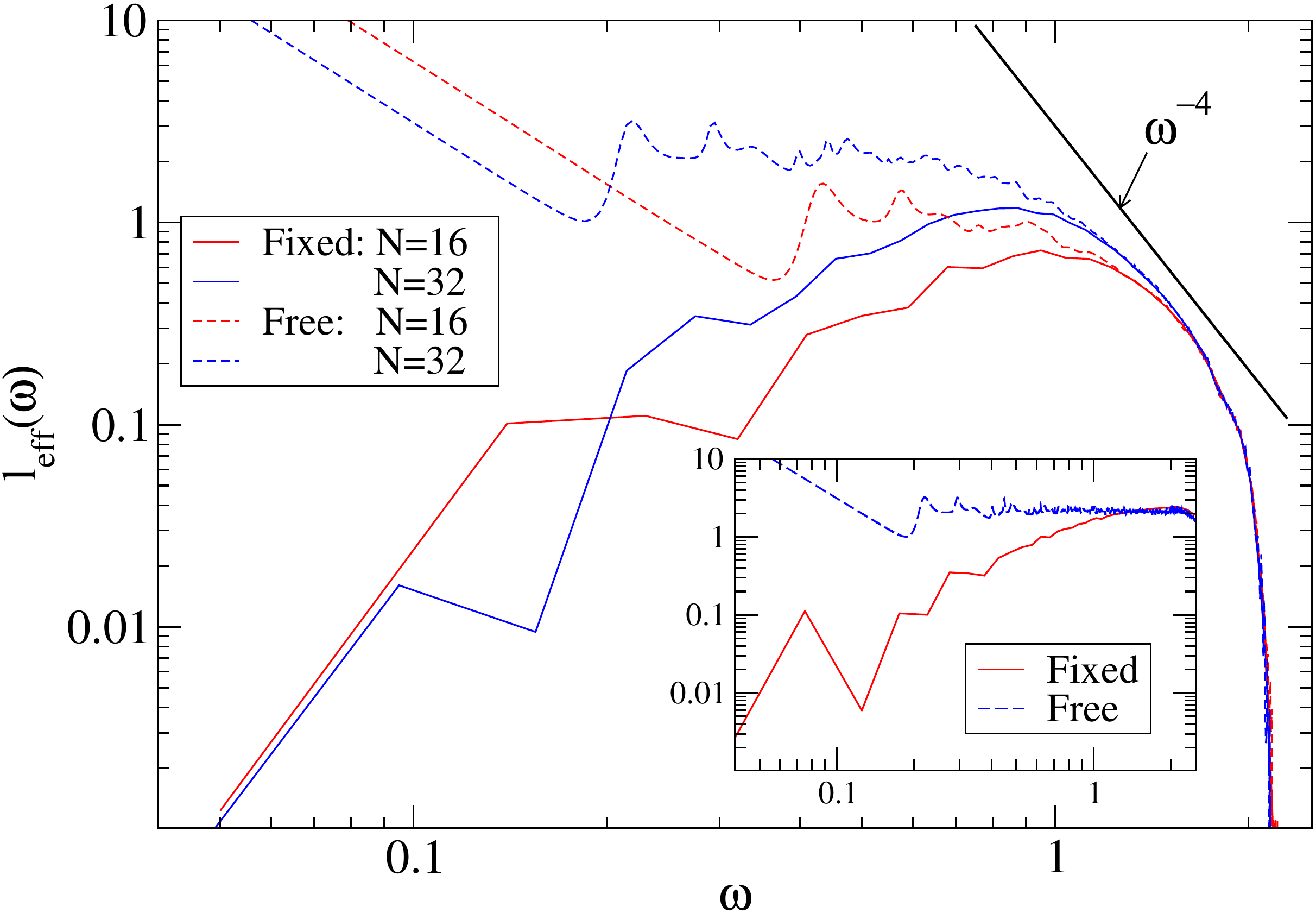}
\label{fig:subfig2}
}
\caption[Optional caption for list of figures]{ 
 Plot of the effective mean-free path $l_{\rm
 eff}=NT(\omega)/\om^{d-1}$ in (a) $2D$ and (b) $3D$ 
 with $\Delta=0.8$. The insets show
 $\ell_{\rm eff}$ for the ordered lattices with a single mass. An 
 $\om^{-(d+1)}$ behaviour is observed in a small part of the diffusive
 region.The fixed BC data is highly oscillatory and has been smoothed. }  
\label{leff}
\end{figure}

For the case of free BCs, we find that the values of  $T(\om)$ in the diffusive
regime matches with those for fixed BCs but are completely
different in the ballistic regime. This is seen in
Fig.~(\ref{leff}) where we plot the
effective mean free path $l_{\rm eff}(\om)=NT(\om)/{\om}^{d-1}$ in the
low-frequency region. The difference between free and fixed BCs is
larger in $2D$ than in $3D$ and explains the similar feature observed
for the $J$ values in Fig.~(\ref{jvsn}). 
For free BC,  $l_{\rm eff}$ is roughly consistent with 
the kinetic theory prediction $l^{-1}_{\rm eff} \sim
N^{-1}+{\ell_K}^{-1}(\om)$ but the behaviour for fixed BC is very
different. The inset of Fig.~(\ref{leff})  plots  
$l_{\rm eff}$ for the equal mass ordered case and we find that in the
ballistic regime it  is very close to the disordered case, an
input that we used in the heuristic derivation.   
The oscillations in the transmission for fixed BC 
arise from scattering and interference of waves at the
interfaces.   
For the fully pinned case  we find $T(\om) \sim e^{-bN}$ in $2D$ and as
${T}(\om)\sim 1/N$ in $3D$ \cite{abhishek09}.

In summary the analytic arguments  show that the
contribution of ballistic modes to conduction is 
dependent on BCs and is strongly suppressed for fixed BCs, the more
realistic case. In $3D$ this leads to diffusive modes dominating for
large system sizes and Fourier's law is satisfied. 
Thus a finite heat conductivity may be obtained without invoking
anharmonicity as is usually believed to be necessary \cite{ziman72}. 
Our numerical results verify the predictions for free BCs and we
believe that much larger system sizes are necessary to verify the fixed
BC results ( this is also the case in $1D$ \cite{dhar01}).
It is also possible that the assumptions made to predict the asymptotic $N$ 
dependence of current is not correct. We are uncertain at present whether the 
disagreements are due to finite size effects or to the inadequacy of the 
theory. A number of recent experiments have directly observed 
localization of electrons \cite{navarro05}, photons \cite{schwartz07},
matter waves \cite{billy08}  and acoustic waves \cite{hu08}. 
Our study shows clearly that the ballistic and diffusive contributions
to the heat current would make it difficult to observe localization
effects in heat conduction studies  \cite{savic08,stoltz08}. 
Our study of the pinned system gives us a clearer understanding of the
role of low frequency modes in giving rise to diverging thermal
conductivity and  provides the simplest example of a $3D$
deterministic system with diffusive transport. The $2D$ pinned system
is a heat insulator and  thin films attached on insulating substrates
may show such behaviour.

We thank G. Baskaran, M. Aizenman,  T. Spencer and
especially D. Huse for  useful discussions. We also thank S. Sastry and
V. Vasisht for use of 
computational facilities.The research of J. L. Lebowitz was supported
by NSF grant No. DMR0802120 and by AFOSR grant No. FA9550-07.

\end{document}